\documentclass[11pt,a4paper]{article}

\usepackage{amsfonts,amssymb,amsthm}
\usepackage{cite} 
\usepackage[T2A]{fontenc}
\usepackage[cp1251]{inputenc}
\usepackage[english]{babel}
\usepackage[english]{babel}
\usepackage{amsmath}%
\usepackage[unicode]{hyperref}
\usepackage[left=1.5cm,right=1.5cm,
    top=2cm,bottom=3cm,bindingoffset=0cm]{geometry}

\allowdisplaybreaks
\numberwithin{equation}{section}

\title{\bf Higher symmetries of the lattices in 3D }
\author{\bf I.T. Habibullin, A.R. Khakimova} 

\begin{document}
\maketitle

\begin{center}
Institute of Mathematics, Ufa Scientific Center, Russian Academy of Sciences\\
 
\end{center}

\abstract {It is known that there is a duality between the Davey--Stewartson type coupled systems and a class of integrable two--dimensional Toda type lattices. More precisely, the coupled systems are generalized symmetries for the lattices and the lattices can be interpreted as dressing chains for the systems. In our recent study we have found a novel lattice which apparently is not related to the known ones by Miura type transformation. In the article we described higher symmetries to this lattice and derived a new coupled system of the DS type.} 

\vspace{0.5cm}


\vspace{0.5cm}

\section{Introduction}

The class of equations of the nonlinear Schr\"odinger type and its spatially two--dimensional analogues is of undoubted interest from the point of view of applications in physics (see, for example, recent works devoted to hydrodynamics \cite{GrinevichSantini} and the dynamics of ferromagnets \cite{Kiselev24}). It is well known that multidimensional integrable models are significantly more complex objects compared to equations of dimension 1+1 and therefore they require the use of fundamentally new ideas and approaches (see \cite{Pogrebkov}, \cite{Adler24}, \cite{BogdanovKonopelchenko}, \cite{Ferapontov}, \cite{Taimanov21}).

It was noted in \cite{ShabatYamilov} that there is a duality between two--dimensional lattices and coupled systems of the Davey--Stewartson type. Namely, coupled systems are generalized symmetries for the lattices. Lattices in turn provide dressing chains for coupled systems. Using this duality, Shabat and Yamilov have found a class of coupled systems corresponding to a known list of lattices containing six models. Among them, they have discovered such an important integrable equation as a spatially two--dimensional generalization of the Heisenberg model. It corresponds to the lattice $E6$ (see the list below). The results of \cite{ShabatYamilov} are important both from the point of view of the integrable classification of both types of equations and from the point of view of finding their explicit particular solutions. The connection of the Davey--Stewartson equation with the Toda chain $E3$ in the context of this duality was noted in \cite{LeznovShabatYamilov}. However, the line of research started in \cite{ShabatYamilov} did not find effective application for a long time due to problems with non-local variables arising in the theory of multidimensional integrable systems (briefly discussed in \cite{LeznovShabatYamilov}). In our recent papers \cite{Habibullin13}, \cite{HabibullinPoptsova18} we showed that integrable discrete and differential-difference equations with three independent variables admit infinite hierarchies of reductions in the form of Darboux-integrable systems of hyperbolic equations of dimension 1+1.
This observation allowed us to partially overcome problems with non-localities both in solving classification problems (see  \cite{HabibullinKuznetsova20}, \cite{Kuznetsova19}, \cite{HabibullinKhakimova21}) and in constructing particular solutions \cite{HabibullinKhakimova24}. It should be noted that an alternative method to the problem of classifying Davey-Stewartson type equations within the framework of the perturbative approach using hydrodynamic reductions is proposed in the work \cite{Novikov}.

Below we explain the aforementioned duality with the second order symmetries of the two--dimensional Volterra chain 
\begin{equation}\label{Volterra}
u_{n,y}=u_n(v_{n+1}-v_n), \qquad v_{n,x}=v_n(u_n-u_{n-1}).
\end{equation}
The simplest generalized symmetry (coupled system) found in \cite{ShabatYamilov} is of the following form
\begin{equation}\label{csVolterra}
\begin{aligned}
&u_{n,t}=u_{n,xx}+\left(u_n^2+2u_nV_n\right)_x,\\
&v_{n,t}=-v_{n,xx}+\left(V^2_n\right)_y+(2u_nv_n)_x, \quad V_{n,y}=v_{n,x}.
\end{aligned}
\end{equation}
It is proven in \cite{ShabatYamilov} that the lattice provides an invertible B\"acklund transformation for  system \eqref{csVolterra}
\begin{equation*}
v_{n-1}=v_{n}-(\ln u_{n-1})_y,\quad u_{n-1}=u_{n}-(\ln v_{n})_x, \quad V_{n-1}=V_{n}-(\ln u_{n})_y.
\end{equation*}
Obviously this transformation just changes value of the discrete argument: $n\rightarrow n-1$.

The Volterra chain admits a large class of symmetries (see \cite{ShabatYamilov}). For instance, one can easily derive another coupled system of the second order from \eqref{csVolterra} by using the involutions $x\leftrightarrow -y$, $t\leftrightarrow \tau$, $u\leftrightarrow v$, $ U\leftrightarrow  V$, $n \leftrightarrow -n$
\begin{equation}
\begin{aligned}
&u_{n,\tau}=u_{n,yy}+\left( U^2_n\right)_x +(2u_nv_n)_y,\\
&v_{n,\tau}=-v_{n,yy}+\left(v^2_n+2v_n U_n\right)_y, \quad  U_{n,x}=u_{n,y}.
\end{aligned}
\end{equation}
The latter admits  the B\"acklund transformation 
\begin{equation*}
u_{n+1}=u_{n}-(\ln v_{n+1})_x, \quad v_{n+1}=v_{n}-(\ln u_{n})_y,\quad U_{n+1}=U_{n}-(\ln v_{n})_y
\end{equation*}
shifting the system forward.
By taking linear combinations of two symmetries given above we find a more complicated symmetry
\begin{equation}\label{csVolterra3}
\begin{aligned}
&u_{n,s}=\lambda u_{n,xx}+\mu u_{n,yy}+\lambda\left(u_n^2+2u_nV_n\right)_x    +\mu\left( U_n^2\right)_x +\mu(2u_nv_n)_y,\\
& V_{n,y}=v_{n,x},\quad  \lambda\neq0,\\
&v_{n,s}= -\lambda v_{n,xx} -\mu v_{n,yy}+\lambda \left(V_n^2\right)_y+\lambda(2u_nv_n)_x +\mu\left(v_n^2+2v_n U_n\right)_y,\\
& U_{n,x}=u_{n,y}, \quad\mu\neq0.
\end{aligned}
\end{equation}

In view of the duality between these two classes, it can be concluded that the presence of a complete list of integrable equations of one of the classes of models would allow obtaining a complete list of integrable representatives of the other class. However, to date, the classification problem has not been solved for either of these two classes.
We have recently made some progress in the problem of classifying lattices. An algorithm for integrable classification of two--dimensional lattices has been proposed, based on the concept of Darboux--integrable finite--field reductions.

The problem of description of the integrable equations of the form
\begin{equation}  \label{todatype}
u_{n,xy} = f(u_{n+1},u_{n},u_{n-1}, u_{n,x},u_{n,y})
\end{equation}
was reduced to the problem of describing all functions $f$ such that hyperbolic systems
\begin{equation}\label{reduc}
\begin{aligned}
&u_{1,xy}=f_1\left(u_1,u_2,u_{1,x},u_{1,y}\right),\\
&u_{j,xy}=f\left(u_{j+1},u_j,u_{j-1},u_{j,x},u_{j,y}\right), \quad 1<j<m, \\
&u_{m,xy}=f_2\left(u_m,u_{m-1},u_{m,x},u_{m,y}\right)
\end{aligned}
\end{equation}
are integrable in the sense of Darboux for arbitrary integer $m\geq2$ for a suitable choice of the functions $f_1=f_1\left(u_1,u_2,u_{1,x},u_{1,y}\right)$ and $f_2=f_2\left(u_m,u_{m-1},u_{m,x},u_{m,y}\right)$. The problem of complete classification in general remains open. But it is solved in the quasilinear case.

For lattices having the following particular quasilinear form
\begin{equation}\label{0}
u_{n,xy}=A_1u_{n,x}u_{n,y}+A_2u_{n,x}+A_3u_{n,y}+A_4    
\end{equation}
where the coefficients depend on the dynamical variables $A_i=A_i(u_{n+1},u_n,u_{n-1})$ for $i=1,2,3,4$, the problem is solved in \cite{HabibullinPoptsova18}, \cite{HabibullinKuznetsova20}, \cite{Kuznetsova19}. Here
we present a list of lattices of class \eqref{0} that passed the test formulated above. This list
is complete up to point transformations.

\begin{itemize}
\item[$(E1)$] $u_{n,xy} = e^{u_{n+1} - 2 u_n + u_{n-1} },$
\item[$(E2)$] $u_{n,xy} = e^{u_{n+1}} - 2 e^{u_n} + e^{u_{n-1}},$
\item[$(E3)$] $u_{n,xy} = e^{u_{n+1}-{u_n}} -  e^{u_n-u_{n-1}},$
\item[$(E4)$] $u_{n,xy} = \left(u_{n+1} - 2 u_n + u_{n-1}  \right) u_{n,x}, $
\item[$(E5)$] $u_{n,xy} = \left(e^{u_{n+1}-{u_n}} -  e^{u_n-u_{n-1}}\right)u_{n,x},$
\item[$(E6)$] $u_{n,xy}=\alpha_nu_{n,x}u_{n,y}, \quad \alpha_n = \frac{1}{u_n - u_{n-1}} - \frac{1}{u_{n+1}-u_n}$,
\item[$(E7)$] $u_{n,xy} = \alpha_n(u_{n,x} - u^2_n - 1)(u_{n,y} - u^2_n - 1) - 2 u_n(u_{n,x}+u_{n,y}-u^2_n - 1).$
\end{itemize}
The list of the coupled systems of the DS type corresponding to the lattices $E1$--$E6$ is given in \cite{ShabatYamilov}.

Note that the system (\ref{Volterra}) considered above does not belong to class (\ref{todatype}), but it is connected with the chain (E5) by a differential substitution (see \cite{ShabatYamilov})
$$u_n=q_{n,x}, \qquad v_n=e^{q_n-q_{n-1}}.$$
In our opinion, this example is simple and clear for the first acquaintance with the discussed approach.

The purpose of this article is studying the novel lattice $E7$
\begin{align}
&u_{n,xy} = \alpha_n(u_{n,x} - u^2_n - 1)(u_{n,y} - u^2_n - 1) - 2 u_n(u_{n,x}+u_{n,y}-u^2_n - 1), \label{newlattice}\\
&\alpha_n = \frac{1}{u_n - u_{n-1}} - \frac{1}{u_{n+1}-u_n}\nonumber
\end{align}
from the generalized symmetries point of view. We notice that equation \eqref{newlattice} is a deep reduction of some more general integrable object defined as a lattice with three--field variables, which may be interesting in its own right. 
Its Lax pair, given by \eqref{Lax7}, is quite unusual. In fact, it depends on the forward/backward difference derivative operators. Based on the ideas of \cite{Ueno}, we developed an algorithm for constructing symmetries for the three--field lattice via this Lax pair. 
Using the algorithm we described hierarchies of symmetries to the three--field lattice. From this hierarchy the desired symmetries for the \eqref{newlattice} are easily found due to the constraint \eqref{reduction}.

\section{Symmetries of an auxiliary three--field lattice}

In this section we concentrate on the problem of constructing generalized symmetries for an auxiliary lattice with three independent variables of the form
\begin{equation}
\begin{aligned}
&a_{n,y}=a_n(b_n-b_{n+1}+u_n-u_{n+1}),\\
&b_{n,x}=b_n(a_{n-1}-a_n+u_{n}-u_{n-1}), \label{gVolterra} \\
&u_{n,y}-a_n(u_{n}-u_{n+1})=u_{n,x}+b_n(u_{n}-u_{n-1})
\end{aligned}
\end{equation} 
with the sought functions $a_n$, $b_n$ and $u_n$. Evidently lattice \eqref{gVolterra} is reduced to the two dimensional  Volterra chain under constraint $u_n=0$. Therefore, it can be considered as a three--field generalization of the Volterra chain \eqref{Volterra}.
Another reduction admitted by \eqref{gVolterra} is given by the relations
\begin{equation}\label{reduction}
a_n=\frac{u_{n,x}-u^2_n-1}{u_{n+1}-u_n}, \qquad b_n=\frac{u_{n,y}-u^2_n-1}{u_n-u_{n-1}}.
\end{equation}
In this case the lattice is transformed into equation \eqref{newlattice}. The third reduction is defined by the constraints
\begin{equation}\label{reduction3}
u_n=0, \quad a_n=\frac{v_{n,x}}{v_{n+1}-v_n}, \qquad b_n=\frac{v_{n,y}}{v_n-v_{n-1}}
\end{equation}
and coincides with equation $E6$
\begin{equation}\label{reductionE6}
v_{n,xy}=v_{n,x}v_{n,y}\left( \frac{1}{v_n - v_{n-1}} - \frac{1}{v_{n+1}-v_n}\right)
\end{equation}
found in  \cite{ShabatYamilov}, \cite{Ferapontov97}. 

The following system of the linear equations (cf. \cite{Kuznetsova21})
\begin{equation}\label{Lax7}
\begin{aligned}
&{\psi}_{n,x}=a_n\left(\psi_{n+1}-\psi_n\right)+u_n\psi_n,\\
&{\psi}_{n,y}=b_n\left(\psi_n-\psi_{n-1}\right)+u_n\psi_n
\end{aligned}
\end{equation}
provides a Lax pair for the lattice \eqref{gVolterra}. In other words this overdetermined system is compatible if and only if  the coefficients $a_n$, $b_n$, $u_n$ solve lattice \eqref{gVolterra}. This fact can be reformulated in terms of the operators
\begin{equation}\label{operators}
\partial_x-B_1 \quad \mbox{and} \quad \partial_y-C_1,
\end{equation}
where $B_1=a_nT-a_n+u_n$ and $C_1=b_n+u_n-b_nT^{-1}$. The shift operator $T$ acts due to the rule $Ty(n)=y(n+1)$. Symbols $\partial_x$ and $\partial_y$  stand for the operators of total differentiation with respect to the variables $x$ and $y$ correspondingly. Actually functions $a_n$, $b_n$, $u_n$ satisfy \eqref{gVolterra} iff the operators \eqref{operators}  commute. 

To construct symmetries of \eqref{gVolterra} we use the method based on the concept of Lax pairs (see, for instance,  \cite{ShabatYamilov}, \cite{Ueno}).
First, we need to describe the class of nonlocal variables on which the symmetries depend. For this purpose, we consider equations
\begin{equation}\label{firstequation}
[\partial_x-B_1, L]=0, \qquad [\partial_y-C_1, M]=0,
\end{equation}
where the sought objects $L$ and $M$ are operators represented as formal power series of the shift operator~$T$
\begin{equation}\label{formalseries}
L=\sum^1_{i=-\infty} \alpha^{(i)}_nT^{i},\qquad M=\sum^{+\infty}_{i=-1} \beta^{(i)}_nT^{i}.
\end{equation}
It is supposed that the first summands of the series are taken as follows
\begin{equation}\label{firscoefficients}
\alpha^{(1)}_n=-a_n,\qquad \beta^{(-1)}_n=b_n.
\end{equation}
The other coefficients are found from equations obtained by comparing the factors in front of the powers of $T$. Actually here we have to solve some linear equations, that generate nonlocalities. For example, the nonlocal variables $H_n:=\alpha^{(0)}_n$ and $V_n:=-\alpha^{(-1)}_n$ are found from equations
\begin{equation}\label{nonlocalitiesL0}
(1-T)H_n=D_x\log a_nb_{n+1}, \qquad D_yH_n=(1-T)a_{n-1}b_n
\end{equation}
and equations
\begin{equation}\label{nonlocalitiesL-1}
(1-T)V_na_{n-1}=D_xH_n, \qquad D_yV_na_{n-1}=D_xa_{n-1}b_n,
\end{equation} 
 respectively.
The coefficients $Q_n:=\beta^{(0)}_n$ and $U_n:=\beta^{(1)}_n$ of the series $M$ are obtained in a similar way
\begin{equation}\label{nonlocalitiesM0}
D_xQ_n=(T-1)a_{n-1}b_{n}, \qquad (T-1)Q_n=D_y\log a_nb_{n+1}
\end{equation}
and, correspondingly, 
\begin{equation}\label{nonlocalitiesM1}
D_xU_nb_{n+1}=-D_ya_nb_{n+1}, \qquad (1-T)b_{n+1}U_n=D_yQ_{n+1}.
\end{equation} 

Thus coefficients of the formal series $L$ and $M$ generate an infinite sequence of the nonlocal variables. It is easy to verify that any positive power of each series satisfies an equation similar to \eqref{firstequation} 
\begin{equation}\label{firstequation-k}
[\partial_x-B_1, L^k]=0, \qquad [\partial_y-C_1, M^k]=0.
\end{equation}
We define new operators $B_k$ and $C_k$ for $k\geq2$ due to the rules
\begin{equation}\label{+-}
B_k=(L^k)_+, \qquad C_k=(M^k)_-.
\end{equation}
Let us explain the meanings of the symbols in \eqref{+-}. Assume that
\begin{equation}\label{formalseries_k}
L^k=\sum^k_{i=-\infty} \alpha^{(i,k)}_nT^{i}\quad \mbox{and} \quad M^k=\sum^{+\infty}_{i=-k} \beta^{(i,k)}_nT^{i}.
\end{equation}
Then we suppose that
\begin{equation}\label{Lk+Mk-}
(L^k)_+=\sum^k_{i=1} \alpha^{(i,k)}_nT^{i}-\sum^k_{i=1} \alpha^{(i,k)}_n,\qquad (M^k)_-=\sum^{-1}_{i=-k} \beta^{(i,k)}_nT^{i}-\sum^{-1}_{i=-k} \beta^{(i,k)}_n.
\end{equation}
It is important that transformations $P_{\pm}$ acting as $P_+:L^k\rightarrow (L^k)_+$ and $P_-:M^k\rightarrow (M^k)_-$ define projection operators.
Note that such methods of truncating formal series differ from the standard methods usually used when searching for higher symmetries of chains in 3D (see, for example, \cite{ShabatYamilov}, \cite{Ueno}).

Such a rule for choosing the polynomial parts of the power series is related to the fact that the basic operators can be written in the following form
$$D_x-B_1=D_x-a_n\Delta_+-u_n,\qquad D_y-C_1=D_y+b_n\Delta_--u_n,$$
where $\Delta_{\pm}=T^{\pm1}-1$ are operators of the forward/backward discrete derivatives. The projection operators $P_{\pm}$ are defined in such a way that the operators $B_k$ and $C_k$ can be represented as follows
\begin{equation}\label{Lk+Mk-2}
B_k=\sum^k_{i=1} \bar\alpha^{(i,k)}_n\Delta_+^{i},\qquad C_k=\sum^k_{i=1} \bar\beta^{(i,k)}_n\Delta_-^{i}. 
\end{equation}
Therefore, all operators involved are polynomials of the  operators $\Delta_+$ or $\Delta_-$.

Since the operators $D_x-B_1$ and $D_y-C_1$ commute with each other, they have to admit common eigenfunctions. In other words, the series $L$ and $M$ can be chosen so that in addition to \eqref{firstequation} the following equations 
\begin{equation}\label{firstequation2}
[\partial_x-B_1, M]=0, \qquad [\partial_y-C_1, L]=0
\end{equation}
are satisfied as well.

Let us take two copies of time flows $x_2, x_3,x_4,\dots,$ $y_2, y_3, y_4, \dots$. Assume that $x_1:=x$, $y_1:=y$. We define a hierarchy of symmetries of the nonlinear system corresponding to these flows by specifying Lax--type representations (cf. \cite{Ueno})
\begin{equation}\label{Lax}
\begin{aligned}
&\partial_{x_k}L=[B_k,L],\qquad   \partial_{x_k}M=[B_k,M],\\
&\partial_{y_k}L=[C_k,L],\qquad   \partial_{y_k}M=[C_k,M].
\end{aligned}
\end{equation}
To search for symmetries, we move, following the method described in \cite{Ueno}, from the Lax--type representation to the Zakharov--Shabat--type equations
\begin{equation}\label{ZakharovShabat}
\begin{aligned}
&\partial_{x_k}B_m-\partial_{x_m}B_k+[B_m,B_k]=0,\\
&\partial_{y_k}C_m-\partial_{y_m}C_k+[C_m,C_k]=0,\\
&\partial_{y_k}B_m-\partial_{x_m}C_k+[B_m,C_k]=0.
\end{aligned}
\end{equation}

{\bf Theorem.} {\it Equations of system 
\eqref{ZakharovShabat} are self consistent, i.e. they produce dynamical systems for arbitrary positive integers $k$ and $m$.}

Proof of the Theorem is given in \S4. It follows from theorem that lattice \eqref{gVolterra} admits an infinite hierarchy of symmetries. The same is true for the lattice $E7$.

Note that equations \eqref{ZakharovShabat} are equivalent to the compatibility conditions of the following set of linear equations
\begin{equation} \label{equationsset}
\Psi_{x_k}=B_k\Psi,\qquad \Psi_{x_m}=B_m\Psi,\qquad \Psi_{y_k}=C_k\Psi,\qquad \Psi_{y_m}=C_m\Psi.
\end{equation}

\section{Examples of symmetries}

\subsection{Searching for the symmetries of the second order}

We use the following two pairs of equations
\begin{equation}\label{ZakharovShabatx2}
\begin{aligned}
&\partial_{x_2}B_1-\partial_{x_1}B_2+[B_1,B_2]=0,\\
&\partial_{x_2}C_1-\partial_{y_1}B_2+[C_1,B_2]=0
\end{aligned}
\end{equation}
and
\begin{equation}\label{ZakharovShabaty2}
\begin{aligned}
&\partial_{y_2}C_1-\partial_{y_1}C_2+[C_1,C_2]=0,\\
&\partial_{y_2}B_1-\partial_{x_1}C_2+[B_1,C_2]=0
\end{aligned}
\end{equation}
to construct the symmetries of the order 2 of \eqref{gVolterra} corresponding to times $x_2$ and $y_2$. 

We define the operator $B_2$ according to \eqref{+-} and \eqref{Lk+Mk-}. As a result of simple calculations we find
\begin{equation}\label{B2}
B_2=a_na_{n+1}T^2-a_n(H_n+H_{n+1})T-a_na_{n+1}+a_n(H_n+H_{n+1}).
\end{equation}

Operator $B_2$ is easily rewritten in terms of the operator $\Delta_+$
\begin{equation}\label{B2delta}
B_2=a_na_{n+1}\Delta_+^2+(2a_na_{n+1}-a_nH_n-a_nH_{n+1})\Delta_++a_na_{n+1}.
\end{equation}

Then we substitute the operators $B_1$, $C_1$ and $B_2$ defined above into system \eqref{ZakharovShabatx2} and after some simplification due to \eqref{gVolterra} and \eqref{nonlocalitiesL0} we arrive at an explicit expression for the desired symmetry  of  system \eqref{gVolterra} in the direction of $x_1$:
\begin{equation}\label{symmfor3}
\begin{aligned}
a_{n,x_2}=&a_{n,x_1x_1}+2a_na_{n,x_1}-2\left(a_nH_n\right)_{x_1}-a_n\left(u_{n,x_1}-u_{n+1,x_1}\right)+a_n\left(u_{n+1}-u_{n}\right)^2\\
&+a_{n+1}a_n\left(u_{n+2}-u_{n+1}\right)-\left(u_{n+1}-u_{n}\right)\left(2a_nH_n-a_n^2-2a_{n,x_1}+a_n\right),\\
b_{n,x_2}=&2b_n\left(a_n-a_{n-1}\right)H_n-b_n\left(a_{n,x_1}+a_{n-1,x_1}\right)-b_n\left(a_n-a_{n-1}\right)^2\\
&-a_nb_n(u_{n+1}-u_n)-a_{n-1}b_n(u_{n}-u_{n-1})+b_n(u_n-u_{n-1}),\\
u_{n,x_2}=&u_{n,x_1}-2a_nH_n\left(u_{n+1}-u_n\right)-\left(u_{n+1}-u_n\right)\left(a_n-a_n^2-a_{n,x_1}\right)\\
&+a_na_{n+1}\left(u_{n+2}-u_{n+1}\right)+a_n\left(u_{n+1}-u_n\right)^2.
\end{aligned}
\end{equation}
To find a symmetry in the $y_1$ direction, we will use the invariance of system \eqref{gVolterra} and the associated non-local variables with respect to the simultaneous replacement of the variables
\begin{equation}\label{involutions}
a\leftrightarrow -b,\quad n\leftrightarrow -n, \quad x_i\leftrightarrow y_i,\quad Q\leftrightarrow H.
\end{equation}
It is clear from this reasoning that the sought symmetry is given by:
\begin{align*}
a_{n,y_2}=&2a_n\left(b_{n+1}-b_n\right)Q_n+a_n\left(b_{n+1,y_1}+b_{y_1}\right)-a_n\left(b_{n+1}-b_n\right)^2\\
&-a_nb_{n+1}(u_{n+1}-u_n)-a_n(u_{n+1}-u_n)-a_nb_n(u_n-u_{n-1}),\\
b_{n,y_2}=&b_{n,y_1y_1}-2b_nb_{n,y_1}-2\left(b_nQ_n\right)_{y_1}-b_n\left(u_{n,y_1}-u_{n-1,y_1}\right)+b_n\left(u_n-u_{n-1}\right)^2\\
&+b_nb_{n-1}\left(u_{n-1}-u_{n-2}\right)+\left(u_n-u_{n-1}\right)\left(2b_nQ_n+b_n^2-2b_{n,y_1}+b_n\right),\\
u_{n,y_2}=&u_{n,y}-2b_nQ_n(u_n-u_{n-1})-(u_n-u_{n-1})\left(b_n^2-b_{n,y_1}+b_n\right)\\
&-b_n\left(u_n-u_{n-1}\right)^2-b_nb_{n-1}\left(u_{n-1}-u_{n-2}\right).
\end{align*}
It is easily obtained from \eqref{symmfor3}.

Using reduction \eqref{reduction} we obtain the symmetries of lattice \eqref{newlattice} in the direction of $x_1$ from the found symmetries of system \eqref{gVolterra}:
\begin{equation}\label{shortx1}
\begin{aligned}
&u_{n,x_2}=u_{n,x_1x_1}-2u_nu_{n,x_1}+u_n^2+1-2(u_n^2-u_{n,x_1}+1)\bar{H}_n, \\
&\bar{H}_n=(T-1)^{-1}D_{x_1}\log \frac{u_{n,x_1}-u^2_n-1}{u_{n+1}-u_n},\\
&D_{y_1}\bar{H}_{n}=- D_{x_1} \frac{u_{n,y_1}-u_nu_{n-1}-1}{u_{n}-u_{n-1}}
\end{aligned}
\end{equation}
and in the direction of $y_1$:
\begin{equation}\label{shorty1}
\begin{aligned}
&u_{n,y_2}=u_{n,y_1y_1}-2u_nu_{n,y_1}+u_n^2+1-2(u_n^2-u_{n,y_1}+1)\bar{Q}_n,\\
&\bar{Q}_n=(T-1)^{-1}D_{y_1}\log \frac{u_{n+1,y_1}-u^2_{n+1}-1}{u_{n+1}-u_{n}},\\
&D_{x_1}\bar{Q}_{n-1}=D_{y_1} \frac{u_{n-1,x_1}-u_{n}u_{n-1}-1}{u_{n}-u_{n-1}}.
\end{aligned}
\end{equation}

Note that the symmetries \eqref{shortx1} and \eqref{shorty1} depend significantly on the discrete parameter $n$, since they contain variables with shifted arguments.
Now our aim is to rewrite them as coupled systems with two unknowns similar to \eqref{csVolterra}. Let us begin with \eqref{shortx1}. Firstly we concentrate on the shifted equation of the form \eqref{shortx1}:
\begin{align*}
u_{n-1,x_2}=u_{n-1,x_1x_1}-2u_{n-1}u_{n-1,x_1}+u_{n-1}^2+1-2(u_{n-1}^2-u_{n-1,x_1}+1)\bar{H}_{n-1}.
\end{align*} 
One can replace the nonlocality $\bar{H}_{n-1}$ due to the relation 
\begin{align*}
\bar{H}_{n-1}=\bar{H}_{n} - D_x\log \frac{u_{n-1,x_1}-u_{n-1}^2-1}{u_{n}-u_{n-1}}. 
\end{align*} 
Afterwards the shifted equation takes the form
\begin{equation}\label{n-1}
\begin{aligned}
u_{n-1,x_2}=&-u_{n-1,x_1x_1}  +2(u_{n-1,x_1}-u_{n-1}^2-1)\bar{H}_{n}-\frac{2u_{n-1,x_1}^2}{u_n-u_{n-1}}\\
&+
\frac{ 2(u_{n-1,x_1}-u_{n-1}^2-1)u_{n,x_1}}  {u_n-u_{n-1}}+
\frac{2(u_{n}u_{n-1}+1)u_{n-1,x_1}}{u_n-u_{n-1}}      +u_{n-1}^2+1,
\end{aligned} 
\end{equation}
where the nonlocality $\bar H_n$ is given by
\begin{align*}
D_{y_1}\bar{H}_{n}=- D_{x_1} \frac{u_{y_1}-uu_{n-1}-1}{u_{n}-u_{n-1}}. 
\end{align*} 
Thus finally we arrive at a coupled system for $u:=u_n$ and $v:=u_{n-1}$:
\begin{equation}\label{csx7}
\begin{aligned}
&u_{x_2}=u_{x_1x_1}-2uu_{x_1}+u^2+1-2(u^2-u_{x_1}+1)\bar{H},\\
&v_{x_2}=-v_{x_1x_1}  +2(v_{x_1}-v^2-1)\bar{H}-\frac{2v_{x_1}^2}{u-v}\\
&\qquad\quad+\frac{ 2(v_{x_1}-v^2-1)u_{x_1}}{u-v}+\frac{2(uv+1)v_{x_1}}{u-v}+v^2+1,\\
&D_{y_1}\bar{H}=- D_{x_1} \frac{u_{y_1}-uv-1}{u-v}.
\end{aligned} 
\end{equation}
Obviously system \eqref{csx7} does not contain any variable with shifted values of $n$.

The second order symmetry of the lattice \eqref{newlattice}  in another direction can also be transformed into a coupled system. To this end we first exclude the variable $\bar Q_n$ according to the formula
\begin{align*}
\bar{Q}_{n}=\bar{Q}_{n-1} - D_y\log \frac{u_{n,y_1}-u_{n}^2-1}{u_{n}-u_{n-1}} 
\end{align*}
and rewrite \eqref{shorty1} as follows
\begin{align*}
u_{n,y_2}=& -u_{n,y_1y_1}+2(u_{n,y_1}-u_n^2-1)\bar{Q}_{n-1}+\frac{2u^2_{n,y_1}}{u_n-u_{n-1}}-\\
&\frac{2(u_{n,y_1}-u_n^2-1)u_{n-1,y_1}}{u_n-u_{n-1}}-\frac{2(u_nu_{n-1}+1)u_{n,y_1}}{u_n-u_{n-1}}+u_n^2+1.
\end{align*}
The nonlocality satisfies the equation
\begin{align*}
D_x\bar{Q}_{n-1}=D_y\log \frac{u_{n-1,x}-u_{n}u_{n-1}-1}{u_{n}-u_{n-1}}. 
\end{align*}
Now we are ready to write down the desired coupled system for the functions $u:=u_n$, $v:=u_{n-1}$
\begin{equation}\label{csy8}
\begin{aligned}
&u_{y_2}= -u_{y_1y_1}+2(u_{y_1}-u^2-1)\bar{R}+\frac{2u^2_{y_1}}{u-v}\\
&\qquad \quad -\frac{2(u_{y_1}-u^2-1)v_{y_1}}{u-v}-\frac{2(uv+1)u_{y_1}}{u-v}+u^2+1,\\
&v_{y_2}=v_{y_1y_1}-2vv_{y_1}+v^2+1+2(v_{y_1}-v^2-1)\bar{R},\\
&D_{x_1}\bar R=D_{y_1}\left(\frac{v_{x_1}-uv-1}{u-v}\right),
\end{aligned} 
\end{equation}
where $\bar{R}_n=\bar{Q}_{n-1}$.

The lattice $E7$, supplemented by the equation for the nonlocality  $\bar{H}_n$, defines the B\"acklund transformation 
\begin{equation}
\begin{aligned}
&v_{n-1}=v_n-\frac{(u_n-v_n)(v^2_n-v_{n,x}+1)(v^2_n-v_{n,y}+1)}{(u_n-v_n)\left(v_{n,xy}-2v_n(v_{n,x}+v_{n,y}-v^2_n-1)\right)+(v^2_n-v_{n,x}+1)(v^2_n-v_{n,y}+1)},\\
&u_{n-1}=v_n,\\
&\bar{H}_{n-1}=\bar{H}_{n} - D_x\log \frac{v_{n,x}-v_{n}^2-1}{u_{n}-v_{n}}
\end{aligned}
\end{equation}
for the coupled system (\ref{csx7}). In a similar way one can derive the B\"acklund transformation for coupled system (\ref{csy8}). Let us give it in an explicit form
\begin{equation}
\begin{aligned}
&u_{n+1}=u_n-\frac{(u_n-v_n)(u^2_n-u_{n,x}+1)(u^2_n-u_{n,y}+1)}{(u_n-v_n)\left(u_{n,xy}-2u_n(u_{n,x}+u_{n,y}-u^2_n-1)\right)+(u^2_n-u_{n,x}+1)(u^2_n-u_{n,y}+1)},\\
&v_{n+1}=u_n,\\
&\bar{R}_{n+1}=\bar{R}_{n} - D_y\log \frac{u_{n,y}-u_{n}^2-1}{u_{n}-v_{n}}.
\end{aligned}
\end{equation}

Note that the lattice (E7) transforms into the lattice (E6) as a result of the replacement of independent variables of the form 
\begin{align*}
x=\varepsilon \bar{x}, \qquad  y=\varepsilon \bar{y},
\end{align*}
with subsequent the limiting transition at $\varepsilon\to 0$. Therefore, the coupled systems for (E7) are transformed into systems for (E6) by means of the substitution 
\begin{align*}
&x_1=\varepsilon \bar{x}_1, \qquad  y_1=\varepsilon \bar{y}_1,\\
&x_2=\varepsilon^2 \bar{x}_2, \qquad  y_1=\varepsilon^2 \bar{y}_2
\end{align*}
using the limit transition. As a result, we obtain the coupled systems for (E6)
\begin{align*}
&u_{\bar{x}_2}=u_{\bar{x}_1\bar{x}_1}+2u_{\bar{x}_1}\tilde{H},\\
&v_{\bar{x}_2}=-v_{\bar{x}_1\bar{x}_1}+2v_{\bar{x}_1}\tilde{H}-\frac{2v_{\bar{x}_1}^2}{u-v}+\frac{ 2v_{\bar{x}_1}u_{\bar{x}_1}}{u-v},\\
&D_{\bar{y}_1}\tilde{H}=- D_{\bar{x}_1} \frac{u_{\bar{y}_1}}{u-v}
\end{align*} 
and
\begin{align*}
&u_{\bar{y}_2}= -u_{\bar{y}_1\bar{y}_1}+2u_{\bar{y}_1}\tilde{R}+\frac{2u^2_{\bar{y}_1}}{u-v}-\frac{2u_{\bar{y}_1}v_{\bar{y}_1}}{u-v},\\
&v_{\bar{y}_2}=v_{\bar{y}_1\bar{y}_1}+2v_{\bar{y}_1}\tilde{R},\\
&D_{\bar{x}_1}\tilde R=D_{\bar{y}_1}\left(\frac{v_{\bar{x}_1}}{u-v}\right).
\end{align*}

\subsection{Searching for the symmetries of order 3}

In this section we construct the third order symmetries of system \eqref{gVolterra}. To this end we use the following two systems of equations:
\begin{equation}\label{ZakharovShabatx3}
\begin{aligned}
&\partial_{x_3}B_1-\partial_{x_1}B_3+[B_1,B_3]=0,\\
&\partial_{x_3}C_1-\partial_{y_1}B_3+[C_1,B_3]=0
\end{aligned}
\end{equation}
and
\begin{equation}\label{ZakharovShabaty3}
\begin{aligned}
&\partial_{y_3}C_1-\partial_{y_1}C_1+[C_1,C_3]=0,\\
&\partial_{y_3}B_1-\partial_{x_1}C_1+[B_1,C_3]=0.
\end{aligned}
\end{equation}

Here operators $B_3$ and $C_3$ are found by virtue of formulas \eqref{+-} and \eqref{Lk+Mk-}. For example, operator $B_3$ has the form:
\begin{equation}\label{B3}
\begin{aligned}
B_3=&-a_na_{n+1}a_{n+2}T^3+a_na_{n+1}(H_n+H_{n+1}+H_{n+2})T^2\\
&-a_n\left(a_{n+1}V_{n+2}+a_{n}V_{n+1}+a_{n-1}V_{n}+H_{n+1}^2+H_n^2+H_{n}H_{n+1}\right)T\\
&+a_na_{n+1}a_{n+2}-a_na_{n+1}(H_n+H_{n+1}+H_{n+2})\\
&+a_n\left(a_{n+1}V_{n+2}+a_{n}V_{n+1}+a_{n-1}V_{n}+H_{n+1}^2+H_n^2+H_{n}H_{n+1}\right).
\end{aligned}
\end{equation}
Similar to the previous case, we substitute explicit expressions of the operators $B_1$, $C_1$ and $B_3$ into system \eqref{ZakharovShabatx3} and obtain an overdetermined system of equations from which we find the symmetry of system \eqref{gVolterra} in the direction of $x_1$:
\begin{align*}
a_{n,x_3}=&-a_{n,x_1x_1x_1}-a_na_{n+1}u_{n+2,x_1}+2a_nu_{n+1,x_1x_1}+a_nu_{n,x_1x_1}+a_n\left(2a_n+3H_n\right)u_{n,x_1}\\
&+3D^2_{x_1}\left(a_nH_n\right)-D_{x_1}\left(a_n^3\right)+3D_{x_1}\left(u_{n+1}-u_n\right)\left(2a_nH_n-a_{n,x_1}\right)\\
&+3D_{x_1}\left(a_n^2H_n-a_na_{n-1}V_n-a_nH^2_n-a_nu_{n+1,x_1}-a_na_{n,x_1}\right)-a_n\left(u_{n+1}-u_n\right)^3\\
&+\left(3a_nH_n-2a_n^2-3a_{n,x_1}\right)\left(u_{n+1}-u_n\right)^2-3a_na_{n+1}\left(u_{n+2}-u_{n+1}\right)\left(u_{n+1}-u_n\right)\\
&-3a_n\left(u_{n+1,x_1}-u_{n,x_1}\right)-a_n\left(a_n^2-3a_nH_n+3H_n^2+3a_{n-1}V_n+7a_{n,x_1}\right)\left(u_{n+1}-u_n\right)\\
&-a_na_{n+1}\left(u_{n+2}-u_{n+1}\right)^2+a_na_{n+1}\left(a_{n+1}+a_{n+2}+3H_n-2a_n\right)\left(u_{n+2}-u_{n+1}\right)\\
&-\left(2a_na_{n+1,x_1}+3a_{n+1}a_{n,x_1}\right)\left(u_{n+2}-u_{n+1}\right)-a_na_{n+1}a_{n+2}\left(u_{n+3}-u_{n+1}\right)\\
&+\left(3a_{n,x_1}-3a_nH_n-2a_n^2+a_na_{n+1}\right)u_{n+1,x_1},\\
b_{n,x_3}=&b_n\left[a_{n,x_1x_1}-a_{n-1,x_1x_1}-3D_{x_1}(a_nH_n)+D_{x_1}\left(a_n\left(u_{n+1}-u_n\right)+a_{n-1}\left(u_{n-1}-u_n\right)\right)\right.\\
&\left.-a_{n-1}\left(u_{n-1}-u_n\right)^2+a_n\left(u_{n+1}-u_n\right)^2-3H_n\left(a_n-a_{n-1}\right)+a_na_{n+1}\left(u_{n+2}-u_{n+1}\right)\right.\\
&\left.+\left(u_{n+1}-u_n\right)\left(a_{n,x_1}+2a_n^2-a_na_{n-1}-3a_nH_n\right)+a_n^3-a_{n-1}^3+3a_na_{n,x_1}\right.\\
&\left.+\left(u_{n-1}-u_n\right)\left(a_{n-1,x_1}+2a_{n-1}^2-3a_na_{n-1}+3a_{n-1}H_n\right)-3a_{n-1,x_1}H_n\right.\\
&\left.+3\left(a_n-a_{n-1}\right)\left(a_{n-1,x_1}-a_na_{n-1}-H_n^2-a_{n-1}V_n\right)\right],\\
u_{n,x_3}=&\left(u_{n+1}-u_n\right)\left(3D_{x_1}\left(a_nH_n\right)-a_{n,x_1x_1}-a_n^3\right)-\left(u_{n+1}-u_n\right)^2\left(2a_n^2-3a_nH_n+2a_{n,x_1}\right)\\
&-3a_n\left(u_{n+1}-u_n\right)\left(H_n^2-a_nH_n+a_{n-1}V_n+a_{n,x_1}\right)-a_na_{n+1}a_{n+2}\left(u_{n+3}-u_{n+2}\right)\\
&-a_n\left(u_{n+1}-u_n\right)\left(u_{n+1,x_1}-u_{n,x_1}\right)-\left(u_{n+2}-u_{n+1}\right)\left(a_na_{n+1,x_1}+2a_{n+1}a_{n,x_1}\right)\\
&-a_n\left(u_{n+1}-u_n\right)^3-a_na_{n+1}\left(u_{n+2}-u_{n+1}\right)\left(2a_n-a_{n+1}-3u_n-3H_n+u_{n+2}\right).
\end{align*}
Due to the invariance of system \eqref{gVolterra} and the nonlocal variables associated with it, according to the following replacement 
\begin{equation}\label{involutions-x3}
a\leftrightarrow - b,\quad n\leftrightarrow -n, \quad x_i\leftrightarrow y_i,\quad Q\leftrightarrow H,\quad U\leftrightarrow V
\end{equation}
we can easily obtain the symmetry of system \eqref{gVolterra} in the $y_1$ direction. We omit these computations.

Finally, by virtue of reduction \eqref{reduction}, we obtain two symmetries of lattice \eqref{newlattice} in the direction of $x_1$
\begin{align*}
&u_{n,x_3}=-u_{n,x_1x_1x_1}+3u_nu_{n,x_1x_1}+3u_{n,x_1}+\left(6u_nu_{n,x_1}-3u_{n,x_1x_1}\right)\bar{H}_n\\
&\qquad \quad-3(u_n^2-u_{n,x_1}+1)\left(\bar{V}_n-\bar{H}_{n,x_1}-\bar{H}_n^2\right)-u_n^2-1,\\
&\bar{V}_n=(T-1)^{-1}D_{x_1} \left(\frac{u_{n,x_1}-u^2_n-1}{u_{n+1}-u_n}-u_n-\bar{H}_n\right),\\
&\bar{H}_n=(T-1)^{-1}D_{x_1}\log \frac{u_{n,x_1}-u^2_n-1}{u_{n+1}-u_n}
\end{align*}
and correspondingly in the direction of $y_1$
\begin{align*}
&u_{n,y_3}=-u_{n,y_1y_1y_1}+3u_nu_{n,y_1y_1}+3u_{n,y_1}+\left(6u_nu_{n,y_1}-3u_{n,y_1y_1}\right)\bar{Q}_n\\
&\qquad \quad-3(u_n^2-u_{n,y_1}+1)\left(\bar{U}_n-\bar{Q}_{n,y_1}-\bar{Q}_n^2\right)-u_n^2-1,\\
&\bar{U}_n=(1-T)^{-1}D_{y_1} \left(\frac{u_{n+1,y_1}-u^2_{n+1}-1}{u_{n+1}-u_{n}}+u_{n+1}+\bar{Q}_{n+1}\right),\\
&\bar{Q}_n=(T-1)^{-1}D_{y_1}\log \frac{u_{n+1,y_1}-u^2_{n+1}-1}{u_{n+1}-u_{n}}.
\end{align*}

\section{Proof of the Theorem}

Here we verify that equations \eqref{ZakharovShabat} are self--consistent and lead to a set of the dynamical systems. At first we concentrate on the case $m=1$, i.e. we examine a pair of the systems
\begin{equation}\label{sym-xk}
\begin{aligned}
&\partial_{x_k}B_1-\partial_{x_1}B_k+[B_1,B_k]=0,\\
&\partial_{x_k}C_1-\partial_{y_1}B_k+[C_1,B_k]=0
\end{aligned}
\end{equation}
and respectively,
\begin{equation}\label{sym-yk}
\begin{aligned}
&\partial_{y_k}C_1-\partial_{y_1}C_k+[C_1,C_k]=0,\\
&\partial_{y_k}B_1-\partial_{x_1}C_k+[B_1,C_k]=0.
\end{aligned}
\end{equation}

Let us rewrite the first equation in \eqref{sym-xk} in the form
\begin{equation}\label{sym-xk-2}
\partial_{x_k}B_1=[\partial_{x_1}-B_1,B_k].
\end{equation}
By construction we have $B_k=L^k-R$ (see \eqref{Lk+Mk-}), where $R$ is given by
\begin{equation}\label{R}
R=\sum^{k}_{i=1}\alpha_n^{(i,k)}+\sum^{i=0}_{-\infty}\alpha^{(i,k)}T^i.
\end{equation}
Equation \eqref{firstequation-k} implies that $\left[\partial_{x_1}-B_1,B_k+R\right]=0$, or the same 
\begin{equation}\label{bar-R}
\bar{R}:=\left[\partial_{x_1}-a_n\Delta_{+}-u_n,B_k\right]=-\left[\partial_{x_1}-a_n\Delta_{+}-u_n,R\right].
\end{equation}
Now we have to examine the relation 
\begin{equation*}
\left[\partial_{x_1}-a_n\Delta_{+}-u_n,\sum^{k}_{i=1}\bar{\alpha}_n^{(i,k)}\Delta^i_{+}\right]=-\left[\partial_{x_1}-a_n\Delta_{+}-u_n,\sum^{k}_{i=1}\alpha_n^{(i,k)}+\sum^{i=0}_{-\infty}\alpha^{(i,k)}T^i\right]
\end{equation*}
to specify $\bar{R}$. The left--hand side implies that $\bar{R}$ may contain a linear combination of positive powers of the operator $\Delta_{+}$ and a free term. On the right we have a free term, the term proportional to $\Delta_{+}$ and negative powers of $T$. Therefore we can conclude that $\bar{R}$ is of the form
\begin{equation}\label{bar-R-final}
\bar{R}=R^{(1)}\Delta_{+}+R^{(0)}, \quad \mbox{where} \quad R^{(0)}=\bar{\alpha}_n^{(1,k)}\left(u_{n+1}-u_n\right).
\end{equation}
Turning back to the relation \eqref{sym-xk-2} we get a couple of equations 
\begin{equation}\label{a-u-xk}
\begin{aligned}
&\partial_{x_k}a_n=-R^{(1)},\\
&\partial_{x_k}u_n=\bar{\alpha}_n^{(1,k)}\left(u_n-u_{n+1}\right),
\end{aligned}
\end{equation}
determining dynamics of the variables $a_n$, $u_n$ in $x_k$.
Now we proceed with the second equation in \eqref{sym-xk}. We rewrite it as follows 
\begin{equation}\label{sym-xk-3}
\partial_{x_k}C_1=[\partial_{y_1}-C_1,B_k].
\end{equation}
As it was remarked above the relation holds $[\partial_{y_1}-C_1,L^k]=0$. Due to formula $L^k=R+B_k$ the latter implies
\begin{equation}\label{bar-S}
[\partial_{y_1}-C_1,B_k]=-[\partial_{y_1}-C_1,R]:=\bar{S}.
\end{equation}
To specify the structure of expression $\bar{S}$ we rewrite relation \eqref{bar-S} in an enlarged form
\begin{equation*}
\left[\partial_{y_1}+b_n\Delta_{-}-u_n,\sum^{k}_{i=1}\bar{\alpha}_n^{(i,k)}\Delta^i_{+}\right]=-\left[\partial_{y_1}+b_n\Delta_{-}-u_n,\sum^{k}_{i=1}\alpha_n^{(i,k)}+\sum^{i=0}_{-\infty}\alpha^{(i,k)}T^i\right].
\end{equation*}
The left side of this relation contains free term, the term proportional to $\Delta_{-}$ and the combination of the positive powers of the operators $\Delta_{+}$. Similarly the right side of the relation contains $\Delta_{-}$, free terms and the negative powers of the operator $T$. Therefore we can conclude that
\begin{equation}\label{bar-S-final}
\bar{S}=S^{(1)}\Delta_{-}+S^{(0)}, \quad  S^{(0)}=\alpha_n^{(1,k)}\left(u_n-u_{n+1}\right).
\end{equation}
Comparing \eqref{sym-xk-3} and \eqref{bar-S-final} we get
\begin{equation}\label{b-u-xk}
\begin{aligned}
&\partial_{x_k}b_n=S^{(1)},\\
&\partial_{x_k}u_n=\alpha_n^{(1,k)}\left(u_{n+1}-u_n\right).
\end{aligned}
\end{equation}
As a result we arrive at the final form of the desired dynamical system
\begin{equation}\label{dyn-sys}
\partial_{x_k}a_n=-R^{(1)}, \qquad \partial_{x_k}b_n=S^{(1)}, \qquad \partial_{x_k}u_n=\alpha_n^{(1,k)}\left(u_{n+1}-u_n\right).
\end{equation}

Now we concentrate on the system \eqref{ZakharovShabat} in the case when $k\geq2$, $m\geq2$. For the definiteness we assume that $m\geq k$. We begin with the first equation in \eqref{ZakharovShabat}. For arbitrary positive integer $s$ we set $R^{(s)}:=L^s-B_s$. Then obviously we have 
\begin{equation}\label{Rs}
R^{(s)}=\sum^s_{i=1} \alpha^{(i,s)}_n  +\sum_{-\infty}^{i=0} \alpha^{(i,s)}_nT^{i}.
\end{equation}
Let us specify the third summand in the equation due to the representation $B_s=L^s-R^{(s)}$. In virtue of the condition $[L^k, L^m]=0$ we arrive at
\begin{equation}\label{mn}
[B_m, B_k]=-[L^m, R^{(k)}]-[R^{(m)},L^k]+[R^{(m)},R^{(k)}]=:S^{(m,k)}.
\end{equation}
On the left of the equation \eqref{mn} we have a polynomial in $\Delta_+=T-1$ of the degree estimated by $m+k$. However, the right-hand side of the expression is a polynomial in $T$ whose degree does not exceed $m$. Consequently the sought function $S^{(m,k)}$ is a polynomial in $\Delta_+$  of the following form 
$$S^{(m,k)}=\sum^m_{i=1} r^{(j)} \Delta_+^j.$$ 
Thus we have the following representation
\begin{equation}\label{eq1}
\partial_{x_k}\left(\sum^m_{i=1} \bar\alpha^{(i,m)}_n\Delta_+^{i}\right)-\partial_{x_m}\left(\sum^k_{i=1} \bar\alpha^{(i,k)}_n\Delta_+^{i}\right)+ \sum^m_{i=1} r^{(j)} \Delta_+^j=0
\end{equation} 
 for the first equation in \eqref{ZakharovShabat}. By comparing coefficients at the powers of the operator $\Delta_+$ we get a dynamical system of the form 
\begin{equation}\label{dynamicalsystem}
\begin{aligned}
&\partial_{x_k} \bar\alpha^{(i,m)}_n- \partial_{x_m} \bar\alpha^{(i,k)}_n + r^{(i)}=0,\quad &&\mbox{for}\quad 1\leq i\leq r,\\
&\partial_{x_k} \bar\alpha^{(i,m)}_n + r^{(i)}=0,\quad &&\mbox{for}\quad r+1\leq i\leq m
\end{aligned}
\end{equation}
generated by the first equation of \eqref{ZakharovShabat}. The second equation in \eqref{ZakharovShabat} is studied in a similar way.

The next step is to study the third equation of \eqref{ZakharovShabat}. For convenience, we write it in the form
\begin{equation}\label{eq3}
\partial_{y_k}B_m-\partial_{x_m}C_k=-[B_m,C_k].
\end{equation}
Due to the representation \eqref{Lk+Mk-2} the l.h.s. of \eqref{eq3} is a linear combination of the positive powers of the operators $\Delta_+$ and $\Delta_-$
\begin{equation}\label{BkCk}
\sum^m_{i=1}\partial_{y_k}\left( \bar\alpha^{(i,m)}_n\right)\Delta_+^{i}-\sum^k_{i=1}\partial_{x_m}\left( \bar\beta^{(i,k)}_n\right)\Delta_-^{i}. 
\end{equation}
Note that \eqref{BkCk} does not contain any free term. Now our aim is to check that the r.h.s. of \eqref{eq3} is of the same form.

It is easily verified that the the following permutation formulas take place
$$\Delta_+\beta_n=\beta_{n+1}\Delta_+ +\Delta_+(\beta_n), \qquad \Delta_-\alpha_n=\alpha_{n-1}\Delta_-+\Delta_-(\alpha_n).$$
For the higher degrees of the operators we have similar relations 
\begin{equation}\label{y3}
\begin{aligned}
&\Delta_+^j\beta_n=\beta_{n+j}\Delta_+^j+r^{(j-1)}\Delta_+^{(j-1)}+\ldots+r^{(1)}\Delta_+ +r^{(0)}, \\
&\Delta_-^i\alpha_n=\alpha_{n-i}\Delta_-^i+s^{(i-1)}\Delta_-^{(i-1)}+\ldots+s^{(1)}\Delta_- +s^{(0)}
\end{aligned}
\end{equation}
with some factors $r^{(p)}$, $s^{(q)}$. They are easily proved by the method of induction. 

The product of powers of the operators $\Delta_+$ and $\Delta_-$ is simplified due to the formula 
\begin{equation}\label{y4}
\Delta_+^i\Delta_-^j=(-1)^i\Delta_+^i+\varepsilon^{(i-1)}\Delta_+^{i-1}+\ldots+\varepsilon^{(1)}\Delta_+ +\varepsilon^{(-1)}\Delta_-+\ldots+\varepsilon^{(-j+1)}\Delta_-^{j-1}+(-1)^j\Delta_-^j
\end{equation}
for $i\geq 1$, $j\geq 1$, here the coefficients $\varepsilon^{(s)}$ are constant integers. We emphasize that the r.h.s. of (\ref{y4}) does not contain any free term.

Let us compute now the commutator of two monomials $\alpha_{n}\Delta_+^j$ and $\beta_{n}\Delta_-^i$:
\begin{align*}
\left[\alpha_{n}\Delta_+^j,\beta_{n}\Delta_-^i\right]&=\alpha_{n}\Delta_+^j\beta_{n}\Delta_-^i-\beta_{n}\Delta_-^i\alpha_{n}\Delta_+^j=\\
&=\alpha_n\left(\beta_{n+j}\Delta_+^j+\ldots+r^{(0)}\right)\Delta_-^i-\beta_{n}\left(\alpha_{n-i}\Delta_-^i+\ldots+s^{(0)}\right)\Delta_+^j.
\end{align*}
Now we replace the products of powers of the operators due to (\ref{y4}) and arrive at the expression 
\begin{equation}\label{y6}
\left[\alpha_{n}\Delta_+^j,\beta_{n}\Delta_-^i\right]=\Sigma^{j=1}_{j'=1}\sigma^{(j')}\Delta_+^{j'}+\Sigma^{i=1}_{i'=1}\delta^{(i')}\Delta_-^{i'}
\end{equation}
that does not contain any free term.

Thus we can conclude that the commutator $\left[B_m,C_k\right]$ is represented in the form
\begin{equation}\label{y7}
\gamma^{(m)}\Delta_+^{m}+\gamma^{(m-1)}\Delta_+^{m-1}+\ldots+\gamma^{(1)}\Delta_+ \gamma^{(-1)}\Delta_+\gamma^{(-2)}\Delta_+^{2}+\ldots+\gamma^{(-k)}\Delta_+^{k}
\end{equation}
with some coefficients $\gamma$. It does not contain terms with $\left(\Delta_+\right)^{0}$ and $\left(\Delta_-\right)^{0}$. Therefore the third equation in \eqref{ZakharovShabat} is self--consistent as well. Theorem is proved.

\end{document}